# TENSORESISTIVE EFFECT IN SINGLE CRYSTAL MICROWIRES OF PbTe DOPED WITH Tl

**Zasavitsky E.A.**

Institute of Electronic Engineering and Industrial Technologies
of  Moldavian Academy of Sciences,
Academiei str.3/3, Chisinau, MD-2028, Moldova
E-mail: efim@lises.asm.md, efim@iieti.asm.md

**Abstract:** Results of room temperature measurements of tensoresistive effect of thin single crystal microwires of $Pb_{1-x}Tl_xTe$ (x=0.0000 ÷ 0.0025, d = 5 ÷ 20 μm) obtained from the melted compound of corresponding composition by the filling of quartz capillary with the following crystallization of material are presented. For the samples corresponding to chemical composition with concentration of thallium x ~0,0025 an essential increase of tensoresistive effect (resistance changes for elastic elongations per unit length of a crystal) in comparison with nondoped samples is observed. Various mechanisms which can lead to observable anomalies, including resonance scattering are discussed. Obtained experimental results allow us to suppose that the observed peculiarities can be interpreted on the basis of model of an impurity band of Tl in PbTe.

## 1. Introduction

Interest to narrow-gap to semiconductor compounds $A^4B^6$ is caused by a lot of unique physical properties and an opportunity of their practical application in various technical devices (infra-red detectors, lasers, tensometers, etc.). In some aspects compounds $A^4B^6$ are unique physical objects - in system PbTe <Tl> superconductivity with superconducting transition temperature being extraordinarily high for semiconductor compounds [1] is observed; in systems PbTe <In> and PbTe <Ga> long-term relaxation processes are observed [2], etc.

The common characteristic property for all these compounds is formation of deep impurity levels and stabilization of a level of Fermi. It is necessary to add, that the impurities stabilizing the Fermi level are also transition elements Cr, Mo and rare-earth elements Yb, Gd [2]. Unique consequence of stabilization of the Fermi level in alloys on the basis of lead telluride is a high homogeneity of electrophysical properties, in spite of strong doping and high concentration of defects [3].

Presence of an impurity band of thallium against the background of the permitted zone states of the valence band of PbTe can lead to essential changes of its electrophysical properties. In particular change of relative position of the impurity band and hole extreme of the valence band under external influence (for example, under influence of deformation) will lead to redistribution of carriers. Change of concentration of carriers in the impurity band (resonant scattering) will be essential to affect electrophysical properties of PbTe doped with thallium. Convenient objects for carrying out of such research are single crystal wires.

Thin single crystal microwires of pure materials possess a number of the peculiar properties making them different from bulk crystals, e.g. elastic properties [4, 5]. Wire crystals of micron thickness admit elastic elongation per unit length up to 2÷5%. Research of influence of uniaxial deformation on electrophysical properties of microwire crystals has been carried out for a lot of materials [4–6].

The scale effect, i.e. substantial increase of mechanical strength of wire crystals at reduction of the sample diameter is explained by reduction of probability of defect occurrence both inside a crystal and on its surface. First of all dislocations sharply decreasing with reduction of the crystal diameter and non-uniform inclusions of impurity pertain to volumetric defects. It being known that the mechanical strength is influenced only by the impurity mechanically captured during growth. The mechanism of reduction of strengthening actions of impurity in wire crystals can be connected with the fact that aggregations of impurity atoms can be a source of nucleation of dislocations.

## 2. Experimental results and discussion

Single crystal microwires of $Pb_{1-x}Tl_xTe$ (diameter d = 5 ÷ 100 μm, length l ~ 20 cm) with thallium average concentration x = 0.00 ÷ 0.02 were grown in the following way [7]. In a quartz tube (diameter - 15 mm) initial material with corresponding chemical composition was placed. The bulk material was prepared in the following way. Since pure Tl oxidizes in the air quickly and greatly, it is necessary for preparation of initial mixtures to use compounds of thallium - in our case TlTe. Syntheses of polycrystalline materials $(PbTe)_{1-x}(TlTe)_x$ of corresponding compounds were made in the quartz tube in the hydrogen atmosphere. Over the material a bunch of quartz capillaries is situated. The choice of quartz as the material for capillaries is limited by the high temperature of its softening, what must be higher than the melting temperature of the material. The tube was evacuated up to residual pressure $10^{-2}$ ÷ $10^{-3}$ Pa and placed in vertical zone furnace, in which the temperature on the whole length of the capillary is the same and higher than the melting temperature of material ($T_{melt}$ < T < $T_{soft}$). After melting of the material the capillaries with open lower ends were put down in the melted material. Afterwards in the tube pressure rose under which capillaries were filled with the melted material. Crystallization of melted material was realized directly beginning from soldered ends to the open ones at the expense of move of the furnace (rate of the move may be changed and makes up several centimeters per hour). The given method of obtaining of single crystal microwires allows producing samples with different diameters under the same growth conditions with high structural perfection. The structural quality was tested by X-ray diffraction and Laser Microprobe Mass Analyzer (LAMMA).

The samples for measurements were prepared in the following mode. The sample of the corresponding diameter was chosen from the set of crystals obtained for carrying out of measurements. As the initial sample has glass isolation, it was preliminarily subjected to selective etching in a solution of acid HF. Measurements were carried out on installation developed in our laboratory.

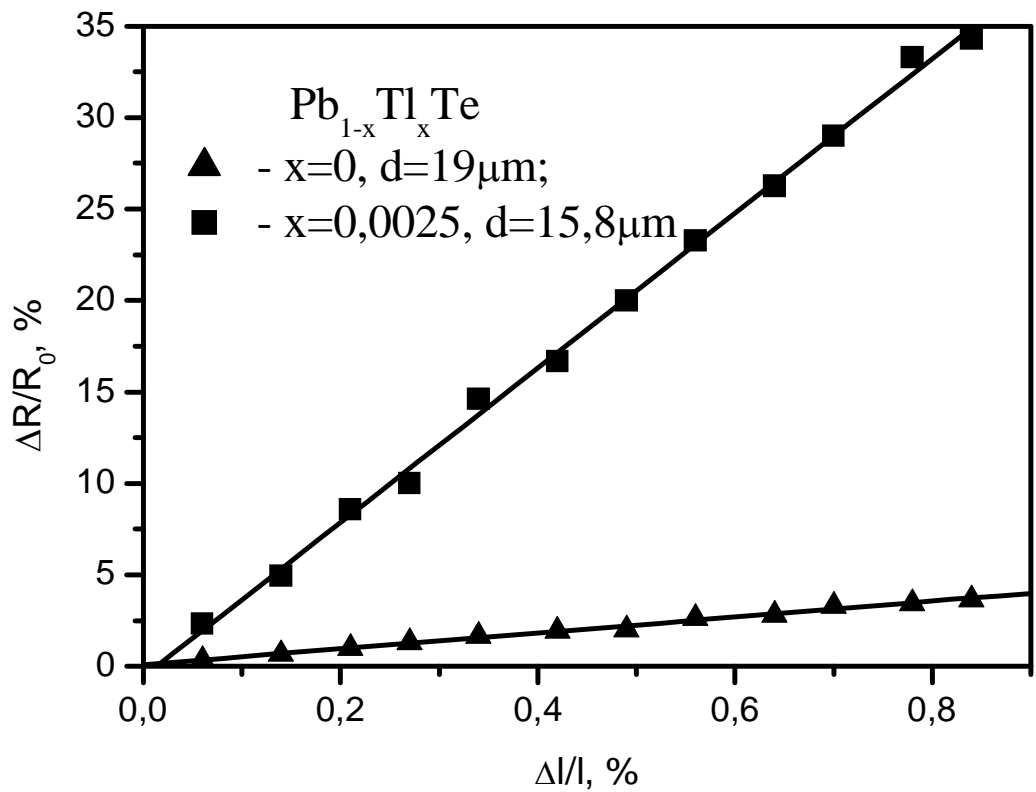

Fig.1. Dependences of resistivity on uniaxial deformation applied to wire monocrystal of $Pb_{1-x}Tl_xTe$.

Fig.1 shows dependence of resistivity on value of uniaxial deformation applied to wire crystal PbTe in a direction similar to the crystallographic direction [100]. The analysis of the dependence shows, that the maximal elastic elongations per unit length of a crystal for all samples achieve the order of 1%. Resistance for all this changes slightly and makes up some percent.

Dependence of resistance on a degree of deformation of doped wire crystal $Pb_{1-x}Tl_xTe$ (x=0,0025) also is presented in Fig.1. One can see that in the case of the doped wire crystals this dependence with other things being equal is by the order stronger.

Uniaxial deformation essentially influences band structure of the material. For films of PbSe and PbTe such researches were carried out earlier [8]. Deformation had elastic character of elongations per unit length down to 6 %. It has been shown, that uniaxial deformation allows realizing semimetal state which cannot be obtained either by means of isotropic deformation or in solid state solutions of type PbTe-SnTe. For all this generally occurrence of semimetal state is connected with various displacements of valleys under action of deformation. In particular, uniaxial deformation applied along the direction [111], leads to a removal of degeneration of four ellipsoids of constant energy of the valence band and of the conduction band in L-point of the Brillouin zone, i.e. so-called valley splitting occurs. Thus one valley along the direction [111] is picked out and other three remain equivalent. It leads to redistribution of charge carriers between valleys that should affect kinetic factors, in particular resistivity of microwire crystals, for example based on PbTe.

The effect of valley splitting is absent in the case of deformation application along the direction [100]. In this case influence of uniaxial deformations on a zone spectrum is reduced only to the change of value of the forbidden zone, and therefore, to the change of value of the effective mass. According to this the piezoresistance effect should be weaker, than in case of deformation along other directions when there is a redistribution of carriers between valleys. Thus the obtained results on research of influence of uniaxial deformations on resistance of pure microwire crystals of PbTe can be well explained within the limits of such approach.

It is considered established, that doping of lead telluride with thallium leads to formation of impurity band against the background of the permitted zone states of the valence band of PbTe [1] (Fig.2).

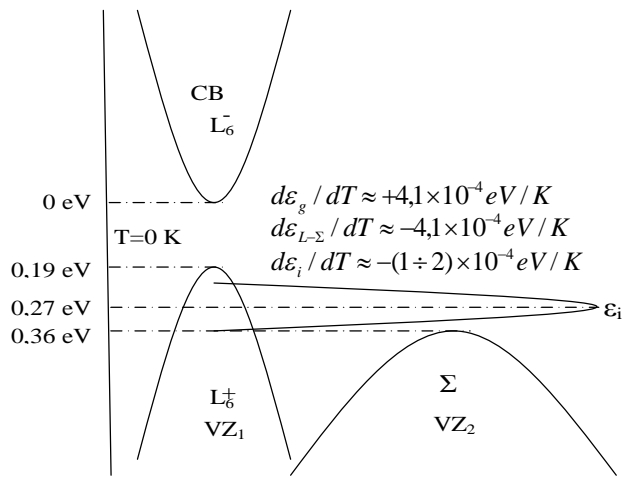

Fig.2 Schematic relative positioning of zones of lead telluride doped with thallium.

Energetic position of a maximum of density of states of thallium impurity band $\varepsilon_i$ at T=77 K makes $\varepsilon_i \approx (0,26 \pm 0,02)$ eV. Thus the width of the impurity band $\Gamma$ of thallium strongly depends on concentration of impurity ($\Gamma = 0,053$ eV at $N_{Tl} = 1,2$ ат.%; $\Gamma = 0,023$ eV at $N_{Tl} = 0,5$ ат.%). With growth of temperature the impurity band is displaced towards edges of the valence band:

$$d\varepsilon_i / dT \approx -(1 \div 2) \times 10^{-4} eV/K$$

Considering, as the gap between extremes of the valence band submits to approximately the same empirical relations:

$$d\varepsilon_{L-\Sigma} / dT \approx -4,1 \times 10^{-4} eV/K,$$

it is possible to state that in a wide temperature region the influence of thallium impurity band on the transport phenomena of lead telluride is actual.

Application of pressure leads to change of the gap between extremes of the valence band according to empirical expression [1]:

$$d\varepsilon_{L-\Sigma} / dP \approx +8 \times 10^{-4} eV/bar,$$

that is, the application of pressure leads to an increase of the gap between heavy-hole band $\Sigma$ and light-hole band L. Thus the pressure dependence of position of the impurity band level is described by formulas [1]:

$$d\varepsilon_i / dP \approx (1 \div 2) meV/kbar.$$

Considering a sign and weak dependence of the impurity level position on pressure, it is possible to draw a conclusion, that the basic contribution to change of electrophysical parameters will be made by movement of the valence band extremes (heavy-hole band $\Sigma$ and light-hole band L).

It is known [1], that density of impurity states $g_i(\varepsilon)$ is described by the Lorenz curve:

$$g_i(\varepsilon) = \frac{N_{Tl}}{\pi} \frac{\Gamma}{(\varepsilon - \varepsilon_i)^2 + (\Gamma/2)^2}.$$

The analysis of this expression allows drawing the following conclusion:
- the strongest dependence of electrophysical parameters will be observed in region of concentration of charge carriers when the Fermi level only starts entering the impurity band;
- in the region of greater concentration of charge carriers in process of the Fermi level approaching to the middle of the impurity band this dependence will become weaker.

Thus in the case of wire crystals $Pb_{1-x}Tl_xTe$ the unique reason of the sharp increase in degree of the resistivity dependence on applied uniaxial deformations alongside with the stated reasons of the resistivity change for wire crystals PbTe, is presence of the impurity band of thallium. In the field of small concentration of thallium impurity when the Fermi level only enters the impurity band, small changes in the concentration of carriers caused by deformation lead to essential change of the mechanism of dispersion of the carriers. Thus the observable features on the resistance dependence on the applied deformation are caused by features of zone of structures PbTe doped with Tl – by the relative energetic position of edges of bands (light-hole band L and heavy-hole band $\Sigma$) and impurity band $\varepsilon_i$.

Investigation of wire crystals $Pb_{1-x}Tl_xTe$ at small impurity concentration is caused by the following reasons. Introduction of an impurity influenced strongly and nonmonotonously the material dislocation structure. Decrease in density of dislocations begins at the certain impurity concentration corresponding to formation of sufficiently powerful impurity atmospheres on dispositions. Experimentally such dependences were observed on many materials - Si, Ge, GaAs [9]. In our case formation of structural complexes on the basis of $Tl_mTe_n$ in the field of small impurity concentration possibly leads to decrease in density of dislocations. It will explain that the doped crystals withstand a load of applied deformations comparable with pure crystals. However at increase in the impurity concentration a partial decay of supersaturated solution PbTe - TlTe is observed. It leads to additional gen-eration of dislocations. It follows from measurements of dependence of the wire crystal microhardness that such boundary concentration of the impurity is equal to 1,125at.%. Besides, crystals with concentration of charge carriers when the Fermi level only starts entering the impurity band are of interest.

## 3. Conclusions

In the present work the results on research of tensoresistive properties of thin single crystal microwires of $Pb_{1-x}Tl_xTe$ (x=0.0000 ÷ 0.0025, d = 5 ÷ 20 μm) at the room temperature for uniaxial deformation down to 1 % are presented. The main task of the given research was detection of influence of thallium impurity on electrophysical properties of lead telluride. It was revealed, that for the samples of PbTe<Tl> corresponding to chemical composition with concentration of thallium x ~0,0025 an essential increase of tensoresistive effect in comparison with nondoped samples is observed. It is qualitatively shown, that this dependence is a function of concentration of charge carriers and should be observed most strongly in the region of concentration where the Fermi level only starts entering the impurity band.

Thus, at presence of the impurity band it is possible to obtain stronger dependences of resistance on deformation, connected with both redistribution of carriers between a zone and an impurity band and with resonant dispersion.

This work was supported by the Supreme Council for Research and Technological Development of Moldova under the State Program "Nanotechnologies, New Multifunctional Materials and Electronic Microsystems".